\documentclass[10pt, conference, compsocconf]{IEEEtran}
\ifCLASSINFOpdf
   \usepackage[pdftex]{graphicx}
\else
\fi
\usepackage[cmex10]{amsmath}

\usepackage{csquotes}

\begin{document}
%
\title{Anomaly Detection for Scenario-based Insider Activities using CGAN Augmented Data}



\author{\IEEEauthorblockN{R G Gayathri, Atul Sajjanhar, Yong Xiang and Xingjun Ma}
	\IEEEauthorblockA{\textit{School of Information Technology} \\
		\textit{Deakin University}\\
		Geelong, VIC 3217, Australia \\
		\{gradhabaigopina, atul.sajjanhar, yong.xiang, daniel.ma\}@deakin.edu.au}}


%


\maketitle

\begin{abstract}
Insider threats are the cyber attacks from the trusted entities within an organization. An insider attack is hard to detect as it may not leave a footprint and potentially cause huge damage to organizations. Anomaly detection is the most common approach for insider threat detection. Lack of real-world data and the skewed class distribution in the datasets makes insider threat analysis an understudied research area. In this paper, we propose a Conditional Generative Adversarial Network (CGAN) to enrich under-represented minority class samples to provide meaningful and diverse data for anomaly detection from the original malicious scenarios. Comprehensive experiments performed on benchmark dataset demonstrates the effectiveness of using CGAN augmented data, and the capability of multi-class anomaly detection for insider activity analysis. Moreover, the method is compared with other existing methods against different parameters and performance metrics.

\end{abstract}


%
\IEEEpeerreviewmaketitle

\section{Introduction}
Organizations are always at high risk from various kinds of cyber attacks. The attacks originated within the organization, referred to as insider attacks, where attackers are in close association with the workplace directly or indirectly and physically or logically are of serious concern. Recently, the reports indicate a drastic rise in the frequency of insider attacks which force the experts to study it more seriously. A study sponsored by ObserveIT and IBM, conducted by Ponemon Institute reveals a 47\% increase in the insider incidents in last two years; 3200 in 2018 to 4700 in 2020 \cite{ponemon}.

Insider threat does not happen in a single day. It needs a lot of time and effort to perform an attack. Various attacking strategies are identified from the past experiences and attack patterns are listed as scenarios \cite{cmu_scenarios}. Detailed study of insider threats reveals that each malicious activity results out of more than one fraudulent action on the trusted resources accessed by the attacker. Earlier research initiatives mapped the malicious insider activities to several scenarios namely Data Exfiltration (Scenario 1), IT Sabotage (Scenario 2), and Intellectual Property Theft (Scenario 3) and so on. 

Though the severity of its impact and the rising numbers and frequency are well-known, the unavailability of real-world data due to privacy reasons and the skewed data distributions marked the insider threat analysis an understudied cyber attack. The insider threat analysis is a typical case of anomaly detection usecase where the anomalous actions are the rarest. Anomaly detection is highly significant in all real-world applications \cite{anomaly}; especially cybersecurity. In general, insider threat analysis is performed as an anomaly detection task, usually as a binary or one-class classification learning problem. 

Supervised learning models built with insufficient data from participating classes tend to overfit thereby learning only the predominant classes resulting in degraded training efficacy. These models fail to generalize on test data. Though many solutions exist to handle overfitting, trying to collect more data is the promising method. Meaningful and reliable data samples representing every single class in the data definitely contribute well to the improved anomaly detection methods. 

To the best of our knowledge, this is the first attempt using a generative model to enrich the minority data samples of various insider activities and perform multi-class classification based anomaly detection. The main aim is to generate data for various underrepresented malicious scenarios that result in a suspicious action. We propose a conditional generative adversarial network (CGAN) to reduce the negative impacts of skewed class distributions in insider threat dataset by boosting the available data and perform anomaly detection. We use t-distributed Stochastic Neighbor Embedding (t-SNE), a manifold-learning based visualization method to perform the qualitative analysis of the CGAN generated data. 

We present a systematic evaluation report that gives the detailed analysis on how the proposed approach improved the performance in terms of popular measures. Comprehensive experiments are performed based on a benchmark dataset under various settings to demonstrate the effectiveness of synthetic data generation for insider detection in comparison with the existing works. 

The rest of this paper is organized as follows. Section \ref{sec_Lit_Review} summarizes the existing insider threat approaches using machine learning (ML) and deep learning (DL) methods and synthetic data generation using CGAN. Section \ref{sec_methodology} presents technical overview of the proposed approach. Section \ref{sec_exp_setup} briefs about the data description being used for validation and the algorithms used followed by evaluation on the performance of our method in terms of various metrics. Section \ref{sec_conclusion} concludes the paper and identifies potential future research directions.

\section{Related Work}\label{sec_Lit_Review}
Insider threat analysis has been studied for many years. But the research community could not contribute a lot in this attack. The limited availability of real-world data has hindered the exhaustive exploration of effective solutions. Nowadays, it attracts wider attention due to the increase in frequency of attacks and the advent of promising data analysis techniques. 

An employee behaving normally is very common; but an insider activity is hard to identify and might go undiscovered. In this situation, the presence of non-malicious activities are significant majority compared to the malicious ones which are very scarce. This kind of highly imbalanced data are unusable as training data for ML/DL models. The significance of class imbalance has not been dealt with clearly and adequately in the literature so far, specifically on AI-based solutions for insider threat analysis.

The recent surveys \cite{insider_survey} reveal that very few publications on insider threat with deep learning have applied methods to balance the disproportionate class samples. In the paper \cite{imbalance_insider1}, oversampling is used to balance the class imbalance followed by an ensemble of one-class classifiers. The authors proposed an Local Outlier Factor (LOF) based oversampling of false-positive instances that are to generate synthetic data points. 

Some other recent works include random oversampling of the training set with various sampling ratios \cite{ieee_trans_scenario} following a binary classification for each scenario, generating manually augmented data using an existing dataset to create an augmented dataset \cite{employee_prof}, spread sub-sampling method \cite{resample_weka} and SMOTE based data adjustment \cite{smote_xgb}.


Recently, data augmentation using vanillaGAN has been performed on insider threat analysis \cite{gan_insider}. The authors used a GAN with fully connected layers to generate the data. Tree-based classifier models are used to test the performance of the data adjusted samples; these models are also prone to overfitting which can result in very high performance scores. The performance seems to be exemplary when compared to all existing models, but the approach makes a lot of assumptions. VanillaGAN convergence and mode collapse, the two tricky aspects of GAN are not dealt with in detail. The qualitative and quantitative evaluation for synthetic data is very vague from the experiments. Moreover, robustness of the data created using generative models is debatable.

The use of adversarial training approaches and synthetic data generation in other application areas like medical imaging and the extensive usage of GANs in cybersecurity problems like intrusion detection motivated us to investigate the potential of GANs for insider threat detection. We discuss the proposed method in detail in the following sections. 
\section{Proposed Method}\label{sec_methodology}
We follow a multi-stage workflow to perform the insider threat analysis. The overall process is split into three sequential stages: (i) Behaviour Extraction (ii) Conditional GAN-based Data Generation (iii) Anomaly detection. First step performs the pre-processing and feature space generation. Second step performs the data augmentation. Finally, the anomaly detection is carried out. Each step is explained in detail in the coming sections.

\begin{figure}[!htpb]
	\centering
	\includegraphics[scale=0.25]{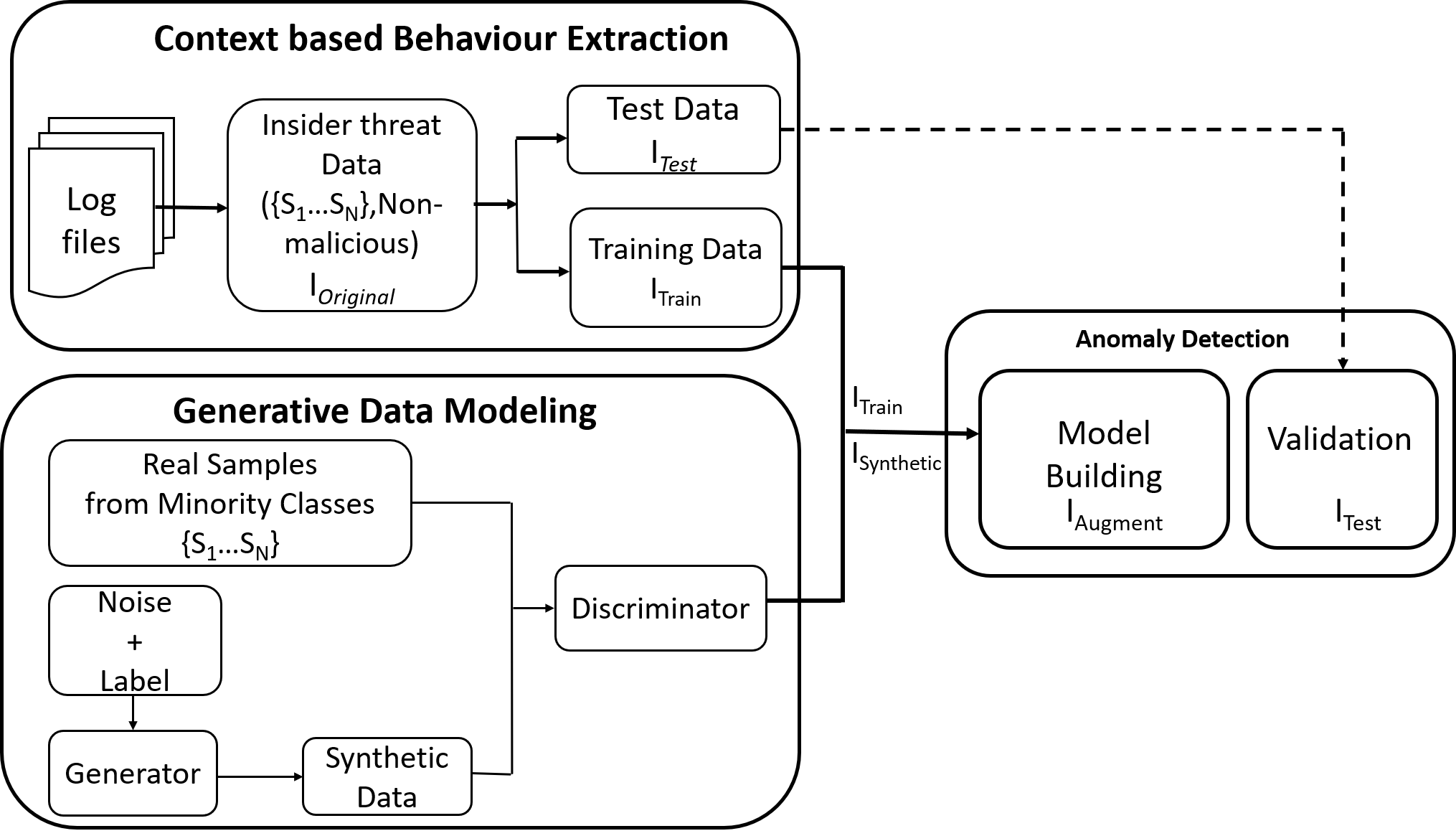}
	\caption[]{Overview of GAN based Insider Threat Analysis}
	\label{fig:overview}
\end{figure}

Let the original dataset be I$_{Original}$ extracted after the feature engineering stage. I$_{Original}$ is split into two subsets: train I$_{Train}$ and test I$_{Test}$. The generative model is used to generate synthetic training data I$_{Synthetic}$. The training split from the original dataset, I$_{Train}$ and the synthetic training dataset, I$_{Synthetic}$ are combined to form the augmented training set I$_{Augment}$ which is used for anomaly detection. Fig. \ref{fig:overview} provides a detailed outline of the proposed workflow. 

\subsection{Behaviour Extraction}\label{sec_ftr_engg}
This section deals with the feature space generation for the insider threat analysis. Context based behavior profiling is used to generate the feature set I$_{Original}$ for the anomaly detection. We used the features from the paper \cite{ieee_trans_scenario}. However, each user is identified as an insider based on the entire activity log instead of those related to the particular scenario. This leads to a context-based user profiling where all the features contribute to the user behavior. The distinct feature sets identified for three scenarios are merged together to form a single feature set to represent context-based features. This makes the feature space easy to scale in cases of new scenario identification.

\begin{figure}[!htbp]
	\centering
	\includegraphics[scale=0.3]{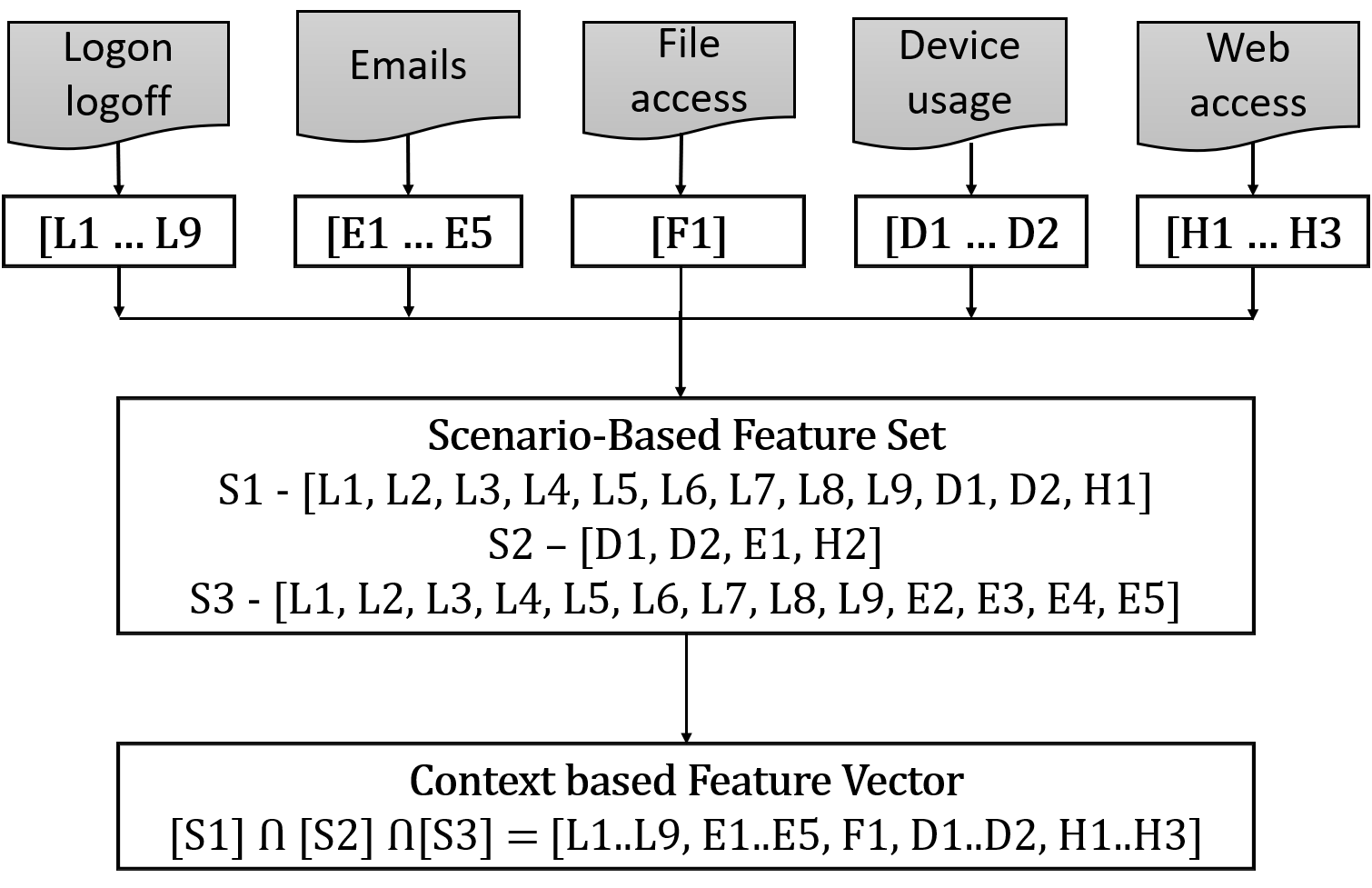}
	\caption{Context Based Daily Feature Set}
	\label{fig:featurespace}
\end{figure}

The various user activities used to represent user behavior are derived from a set of log files. The user activity logs obtained from the logon-logoff details of the users, file access history, external device usage patterns, email communications and the web browsing history files are pre-processed as per the explanation provided in \cite{ieee_trans_scenario}. The feature engineering detailed in this section is not restricted to any particular dataset; but can be used for any data with minimal adaptation depending on the availability of the log files. The summary of feature space generation is depicted in Fig. \ref{fig:featurespace}. 

The difference between the feature set generation of the proposed approach and the one used in the work \cite{ieee_trans_scenario} is in the two features H2 and H3; those depicting web browsing history. H1, a frequency-based feature, is computed as the number of times a user visits the website wikileaks.org. The previous work followed a TF-IDF based feature extraction creating a corpus with the data available within the log files provided as part of the dataset. But this method seems to be unsuitable as the corpus generation is done using the keywords provided by the data itself; the corpus not being exhaustive.

We propose a Jaccard similarity index based score to compute H2 and H3. Jaccard similarity coefficient gives a score between 0 and 1 where 0 indicates no similarity and 1 indicates a similarity. This can be interpreted as the correlation between related keywords. Jaccard Similarity based methods are used in information retrieval systems and text processing. 

In the proposed method, we build two corpora D1 and D2. D1 consists of terms related to job opportunities; keyword list from Scenario 1. D2 contains terms, such as keylogger, password, crack etc; keyword list from Scenario 2 and Scenario 3. Jaccard Similarity Coefficient is computed for each set of keywords from the individual user actions to get the features H2 and H3. 
Let us consider two sets of keywords X and Y where X is the keyword list for the daily activity and Y is the keyword list in the corpus. The computation of H1 and H2 features are as follows:

\begin{equation}\label{eq_jdindex}	
	\begin{split}
		H2 = J(keywords,D1) \\	
		H3 = J(keywords,D2) \\
		J(X,Y) = \frac{|X \cap Y|}{|X\cup Y|}
	\end{split}
\end{equation}

The pre-processed dataset I$_{Original}$ consists of various insider activities; hence classes are labeled on the individual scenarios. The feature set comprehensively describe the daily resource usage pattern of each user which results in an extremely imbalanced class distribution. Many works have been done on balancing the data using re-sampling methods, cost-sensitive approaches, etc. In this work, our main aim is to use a generative model to balance the data distribution and build a supervised learning model for anomaly detection. Next section explains how generative adversarial networks are utilized to reduce the skewed class distribution.

\subsection{Conditional GAN-based Data Augmentation}\label{sec_gen_syn_data}
One of the main limitations that restricts the insider threat analysis is the lack of real-world datasets. Generative Adversarial Networks (GAN) \cite{gan}, a variant of neural networks, triggered a trusted way to generate synthetic data with the same properties as of a given data distribution. GANs are widely used in image processing to create visually similar images as that of the real images such that an adversary finds it difficult to distinguish them. Though GANs are popular for data formats like text and images, their adoption in tabular data with numerical data types are not that common at present. Our feature set comprises of numeric attributes; therefore we design a new CGAN network which is less complex and suitable for the tabular 2D data.

We propose a conditional-GAN (CGAN) which is conditioned by the minority malicious class labels. The two-folded benefit of using the CGAN generated data are : (i) It helps to reduce the impact of class imbalance (ii) The CGAN helps in increasing the diversity of training set by including data from all classes. Conditional GAN generates data conditioned on class labels via label embeddings in both discriminator and generator. The conditional GAN worked well in all settings and proved to be acceptable for insider threat analysis. In the network, the discriminator (D) tries to distinguish whether the data is from the real distribution, while the generator (G) generates synthetic data and tries to fool the discriminator. We use a fully connected neural network in the generator and discriminator. The CGAN network architecture is as shown in Table. \ref{tab:arch}.

\begin{table}[!hbtp]
	\caption{Proposed CGAN Network Architecture}
	\label{tab:arch}
	\begin{tabular}{lccc}
		\hline
		\textbf{Operations} & \textbf{Units}                          & \textbf{Non Linearity} & \textbf{Dropout} \\ \hline
		\multicolumn{4}{l}{\textit{Generator}}                                                                    \\ \hline
		Dense               & 32                                      & LeakyReLU              &                  \\
		Dense               & 64                                      & LeakyReLU              &                  \\
		Dense               & 128                                     & Linear                 &                  \\ \hline
		\multicolumn{4}{l}{\textit{Discriminator}}                                                                \\ \hline
		Dense               & 256                                     & LeakyReLU              &                  \\
		Dense               & 128                                     & LeakyReLU              & 0.2              \\
		Dense               & 32                                      & LeakyReLU              &                  \\
		Dense               & 1                                       & Sigmoid                &                  \\ \hline
		\multicolumn{4}{l}{\textit{CGAN Parameters}}                                                               \\ \hline
		Optimizer           & \multicolumn{2}{l}{Adam (lr=0.0002, beta\_1=0.5)}                 &                  \\
		Epochs              & \multicolumn{1}{l}{300}                 &                        &                  \\
		Batch Size          & \multicolumn{1}{l}{64}                  &                        &                  \\
		Conditioned         & \multicolumn{1}{l}{On Minority Classes} &                        &                  \\
		Latent Dimension    & \multicolumn{1}{l}{20}                  &                        &                  \\ \hline
	\end{tabular}
\end{table}

The design is made out of a simple configuration for generating usable synthetic numerical data for insider threat. The generator and discriminator are fully connected layers with \{32, 64 and 128\} and \{256, 128 and 32\} neurons respectively. Latent dimension input to the generator is same as the number of features. The discriminator is regularized using a drop out of 0.2 in the penultimate layer. Adam optimizer, Leaky ReLU and dropout regularization are used as they are the standard followed in similar problems. Binary cross-entropy is used as the adversarial loss function as it suits well for a model performance with an output probability between 0 and 1. The training involved 300 epochs. Training more than the given epochs did not provide any improvement in the generator performance. 

There is an extremely low sample count for certain classes. These classes are left unpicked during random selection of original samples in the training phase. Therefore, in the network design, the random real samples are chosen such that all the classes are given equal importance in the CGAN training. This ensures that no class is left behind in any of the iterations such that discriminator gets all classes in each step. Any number of samples from any minority class can be generated by providing the type of class required as input. The new samples are merged into the original training distribution. As shown in Fig \ref{fig:overview}, the synthetic dataset I$_{Synthetic}$ is combined with I$_{Train}$, the training dataset from I$_{Original}$. We provide test results and comparisons in the coming sections to support the proposed data augmentation step for insider threat datasets. 

\subsection{Anomaly Detection - Multi Class Classification}\label{sec_anomaly_det}
Anomaly detection using ML/DL algorithms are popularly accepted approach for insider threat detection. Existing approaches had tended to focus on supervised learning using two classes : Malicious and Non-malicious. One of the major reasons for the binary classification based analysis is the lack of insider threat datasets and the diversity of the malicious activities. Tree-based models are known to perform well on small data whereas deep learning models need significant number of samples in its training dataset to learn the representations to build a robust classifier. Though there are algorithms like XGBoost which perform well on the imbalanced data, the need for other complex learning models like artificial neural networks prevail. We chose a mix of ensemble algorithms and artificial neural networks which are gaining popularity recently.

Binary classification is a special case of multi-class classification where each malicious class is modeled as a separate classification problem. In other words, it follows a multiple binary classification approach or One-vs-Rest strategy. In this method, one builds independent classifiers for each class. None of the existing methods tried to merge the results to a multi-class set up. On the other hand, a multinomial classifier learns directly from the available classes. This means that the parameters are estimated in a class interdependent manner thereby building models that are robust against outliers. 


Unlike existing methods, the proposed method considers the pre-defined scenario based malicious activities and try to give a deeper insight into the employee activities. Anomaly detection considers the different scenarios under malicious class and normal activities as non-malicious instances and perform the multi-class classification. The classifier tries to discriminate the anomalous activities from rest of the samples. In the paper \cite{imbal_survey}, multi-class classification and its importance for learning from imbalanced data are dealt with in detail. The work suggests the effectiveness of neural networks and ensemble based methods for imbalanced data handling.


The algorithms chosen for supervised learning include XGBoost, Random Forest (RF), Multilayer Perceptron (MLP) and 1-Dimensional Convolutional Neural Network (1DCNN). XGBoost, a tree-based ensemble algorithm, proved to be very efficient for all kinds of data. Hence, many existing works on insider threat analysis use XGBoost for modeling \cite{ieee_trans_granular}. Random Forest, yet another tree-based ensemble algorithm, is also a popular one for imbalanced data. We chose MLP and 1DCNN as the neural network algorithms for our experiment. MLP, a popular neural network, is suitable for numerical data and known for better generalization. 1DCNN, a CNN model that operates over 1D sequences, popular for time-series data is gaining acceptance in other domains like intrusion detection \cite{1dcnn_intrusion}.

\section{Experiments}\label{sec_exp_setup}

In this section, we evaluate the performance of the proposed method. We provide the dataset description followed by the models being used for training and performance evaluation. We have implemented the method using Python programming language, Tensorflow and Keras. 

\subsection{Dataset Description}
Real-world data scarcity caused a major hindrance to the experimentation. Usual practice is to use either data collected from real user data, or use synthetically generated data. Malicious insiders are predominantly employees in an organization with access rights. Data collection involves direct tracking and monitoring of the user actions and behaviors of the employees. This creates privacy and confidentiality concerns in an organization. Therefore, in cases like this researchers are forced to progress with synthetic data. 

We used the CMU CERT dataset \cite{cmu}, the widely accepted synthetic data for insider threat detection. There are various versions of CERT datasets available. CERT v4.2 is commonly used as it has the maximum cases of insiders grouped into three scenarios. The data spans over 500 days with 1000 users and 70 insiders. Table \ref{tab:dataset} represents the summary of activity logs used for pre-processing.

\begin{table}[!hbtp]
	\caption{Summary of the heterogeneous log files}
	\label{tab:dataset}
	\centering
	\begin{tabular}{|c|c|}
		\hline
		Data & No.of Instances \\
		\hline
		Logon.csv & 854859 \\
		Email.csv & 2629979 \\
		Http.csv & 1048575 \\
		Device.csv & 405380 \\
		File.csv & 445581 \\
		\hline
		\hline
		Pre-processed Data Summary & \\
		\hline
		Non-malicious events & 7154815\\
		Malicious events	& 7323 \\
		No. of features & 20\\
		Total instances & 330452\\
		Malicious & 966 \\
		Non-Malicious & 329487\\
		
		\hline
	\end{tabular}
\end{table}

Synthetic data comprises of various log files as depicted in Fig \ref{fig:featurespace}. Pre-processed data consists of 330452 non-malicious instances and 966 malicious instances. The pre-processed data are labeled into four classes namely non-malicious, S1, S2 and S3. It is obvious that the data distribution is extremely skewed, where malicious insider related data counts to only 1.278\%. 
Scenario 3 (IT sabotage related malicious actions) has the least amount of users and data instances. Maximum number of instances are for Scenario 2 followed by Scenario 1.

\subsection{Authenticity of Synthetic Data}

Metrics for measuring the similarity of synthetic data against original data distribution for numerical data types are not well established. In case of images, there are metrics like Inception Score to compare the synthetic and real images. Due to the unavailability of any proven metric to compare the numerical data similarity between the original and synthetic data distribution, we use visual approach based on kernel density estimates (KDE), the Principal Component Analysis (PCA) and t-distributed Stochastic Neighbor Embedding (t-SNE) on the original and synthetic data. 

\begin{figure}[!htbp]
	\centering
	\includegraphics[scale=0.3]{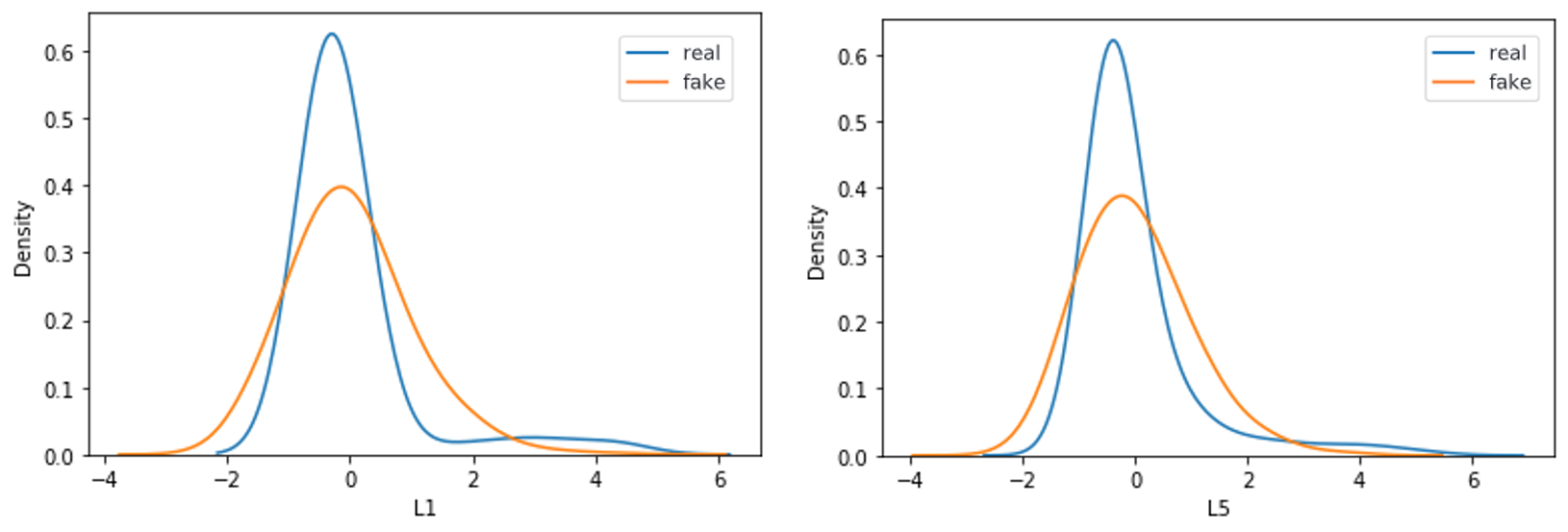}
	\caption{Kernel density estimation of original and synthetic data}
	\label{fig:kde}
\end{figure}

We used a kernel density estimation, a non-parametric method, on the original and synthetic data to show the data distribution similarity. Due to space constraint, we depict two features using the visualization. Fig. \ref{fig:kde} depicts the kernel density estimation plot on two features L1 and L5.

\begin{figure}[tbph!]
	\centering
	\includegraphics[scale=0.27]{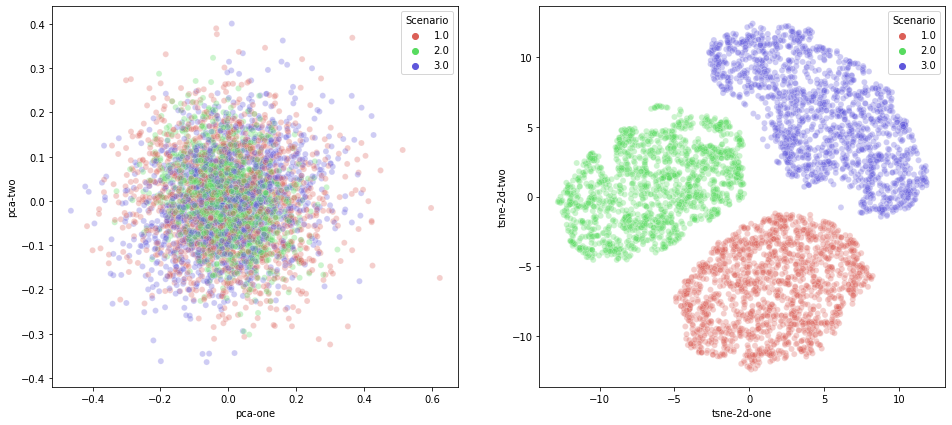}
	\caption{PCA and t-SNE based visualization}
	\label{fig:pcatsnefake}
\end{figure}

PCA is a technique that converts n-dimensions of data into k-dimensions while maintaining as much information from the original dataset. PCA differs from t-SNE where the later one preserves the local neighbors of the data points. t-SNE can be considered as a manifold learning where the geometric data properties are used. t-SNE helps to increase the interpretability of data in the lower dimensions. Fig. \ref{fig:pcatsnefake} shows the PCA and t-SNE visualization of the data on 2D. PCA shows the dimension reduced data. Since it is a 2D visualization, the separate clusters are not visible; hence the overlap. In the t-SNE based visualization, the three clusters are the minority insider activity samples generated using the CGAN model. The larger spread of the data show that there is no evident mode-collapse.

\subsection{Performance Metrics}

The classification models are validated against common performance metrics for imbalanced data like precision (P), recall (R) and f-score(F). It is highly recommended to have high precision and recall. Sometimes these measures alone do not reflect the true performance of various models. Usually, the majority class is considered as the negative class. It is because of the assumption that interesting samples are rare and hence, considered as positive. 

Though the existing works achieved a high precision and recall, no work has performed a detailed analysis of learning reports. In addition to analyzing the results based on precision and recall, we validated the results against the number of false-positives and false-negatives. Considering the prime importance of the confusion matrix, also referred to as an error matrix, we have included two more metrics, namely, Cohen's Kappa (Kappa) and Mathews Correlation Coefficient (MCC) \cite{mcc} for the experimental evaluation.

Cohen's Kappa or Kappa score measures the degree of agreement between the true values and the predicted values. Cohen’s kappa is always less than or equal to 1. Values of 0 or less, indicate that the classifier is useless. A value between 0.81–1 is interpreted as almost perfect agreement; the score reaches its maximum for balanced data. 

Precision, recall and F-score are asymmetric; when the classes are interchanged, their values change. Precision and recall consider the positive class to be the class of interest. True-Positive (TP), False-positive (FP), and False-Negative (FN) from the confusion matrix are used for their computation. True-Negative (TN) — is not used in these metrics. Changes in TN never reflect the precision and recall. Hence, we consider MCC, a symmetric metric which gives value between -1 and +1 inclusive. MCC considers all four values in the confusion matrix; no class is more important so that switching the negative and positive classes will give same value. A value close to 1 means that both classes are predicted well. 

\subsection{Results and Discussion}
In this section, we explain the performance of multi-class anomaly detection approach using tree-based and ANN algorithms on insider detection. Comprehensive set of experiments were conducted on original data as well as applying various data augmentation methods as follows : (i) on original skewed data with no data augmentation (ii) on Random Over Sampled (ROS) training data (iii) on SMOTE oversampled training data and (iv) GAN generated synthetic data in the training set. Table \ref{tab:multiclass} gives the performance of anomaly detection using multi-class classification.

\begin{table}[!htbp]
	\centering
	\caption{Performance Analysis of Multi-class Classification}
	\label{tab:multiclass}
	\begin{tabular}{|l|l|l|l|l|l|}
		\hline
		\textbf{Training Models} &
		\multicolumn{1}{c|}{\textbf{P}} &
		\multicolumn{1}{c|}{\textbf{R}} &
		\multicolumn{1}{c|}{\textbf{F}} &
		\multicolumn{1}{c|}{\textbf{Kappa}} &
		\textbf{MCC} \\ \hline
		Real + RF       & 0.482          & 0.645          & 0.482          & 0.465          & 0.512 \\ \hline
		ROS + RF        & 0.438          & 0.658          & 0.479          & 0.347          & 0.431                      \\ \hline
		SMOTE + RF      & 0.596          & 0.735          & 0.648          & 0.740          & 0.746                      \\ \hline
		\textbf{CGAN + RF}  & \textbf{0.751}          &\textbf{0.748}          & \textbf{0.739}          & \textbf{0.920}         & \textbf{0.920}      \\ \hline	  \hline
		Real + XGBoost   & {0.749}       & {0.655}        & {0.697}        &  {0.877}       & {0.884}             \\ \hline
		ROS + XGBoost   & 0.368          & 0.866          & 0.419          & 0.156          & 0.290                      \\ \hline
		SMOTE + XGBoost & 0.396          & 0.771          & 0.431          & 0.120          & 0.250                      \\ \hline
		\textbf{CGAN + XGBoost}        & \textbf{0.782} & \textbf{0.761} & \textbf{0.759} & \textbf{0.891} & \textbf{0.886}             \\ \hline \hline
		Real + MLP      & 0.275          & 0.788          & 0.283          & 0.055          & 0.168                      \\ \hline
		ROS + MLP       & 0.406          & 0.790          & 0.457          & 0.162          & 0.296                      \\ \hline
		SMOTE + MLP     & 0.452          & 0.774          & 0.492          & 0.156          & 0.289                      \\ \hline
		\textbf{CGAN + MLP}       & \textbf{0.833} & \textbf{0.767} & \textbf{0.733} & \textbf{0.857} & \textbf{0.805}             \\ \hline  \hline	
		Real + 1DCNN    & 0.267          & 0.932          & 0.263          & 0.036          & 0.136                      \\ \hline
		ROS + 1DCNN     & 0.536          & 0.793          & 0.531          & 0.245          & 0.371                      \\ \hline
		SMOTE + 1DCNN   & 0.527          & 0.778          & 0.524          & 0.233          & 0.360                      \\ \hline
		\textbf{CGAN + 1DCNN}     & \textbf{0.799} & \textbf{0.767} & \textbf{0.763} & \textbf{0.821} & \textbf{0.831}             \\ \hline
	\end{tabular}
\end{table}

ML/DL methods were used for building anomaly detection models on all the above mentioned data settings. We use text highlighted in bold to depict the interesting results in the tables with experimental results. The tables use the abbreviations Precision (P), Recall (R), F-score (F), Cohen's Kappa (Kappa) and Mathews Correlation Coefficient (MCC). 

The results from RF algorithm is disappointing in the multi-class classification on the original data; but it gave noticable increase in performance on using more meaningful data for training. Compared to ROS and SMOTE, CGAN shows remarkable improvement in all performance metrics. In case of RF, there are chances of overfitting with ROS and SMOTE data augmentation. In order to reduce the effect of overfitting, we used the gradient boosting ensemble (XGBoost). The ROS and SMOTE oversampling do not show any improvement in XGBoost whereas the real data and the CGAN boosted training data resulted in a better model creation. With CGAN, XGBoost shows reasonable rise in the precision and f-score metrics. 

The deep learning algorithms need more data to learn from the deep representative features. Hence, MLP and 1DCNN show remarkable improvement in the metrics. We also used the Cohen's Kappa measure and MCC to validate the performance. The CGAN enriched dataset yielded the strong performance in terms of Kappa score. MCC, a correlation value shows a significant improvement for the CGAN enriched multi-class classification experiments. The performance of GAN augmented data for the anomaly detection is evident from the striking improvement in the metrics as shown in Table \ref{tab:multiclass}. The results offer compelling evidence for supporting the efficiency of our proposed method.

\subsection{Comparison with Existing Methods}
In this section, we discuss experimental results of binary classification on the proposed method using same features. We conducted scenario-based binary classification and the results are illustrated in Tables \ref{tab:compares1}, \ref{tab:compares2} and \ref{tab:compares3} using the feature space generation in Section \ref{sec_ftr_engg}.

The same CGAN architecture is used to generate samples for each class. For instance, to perform the binary classification for Scenario 1 vs non-malicious samples, we generated data for scenario S1 using the CGAN model and considered all other samples as non-malicious class. 

We present the binary classification on three sets of training data : (i) original data (ii) SMOTE oversampled data and (iii) original data boosted with CGAN generated synthetic data. In case of Scenario 1, the original data has 85 instances which is significantly low in number for complex models to learn the patterns from a high volume of negative samples. The results show that XGBoost could perform moderately for the original data. But the precision is low with a higher number of false-positives. Though the SMOTE oversampling did not make any improvement, the synthetic data from GAN merged with original training data gives a remarkable decrease in the number of false-positives for all models which is evident in all metrics. 
Table \ref{tab:compares1} gives the analysis of the insider activities using the proposed method with scenario 1 and all other cases being non-malicious. 
\begin{table}[!htbp]
	\centering
	\caption{Performance analysis of Scenario 1 Binary Classification }
	\label{tab:compares1}
	\begin{tabular}{|l|l|l|l|l|l}
		\hline
		\multicolumn{6}{|c|}{\textbf{Scenario 1}}\\ \hline
		\textit{\textbf{Training Models}} &
		\multicolumn{1}{c|}{\textit{\textbf{P}}} &
		\multicolumn{1}{c|}{\textit{\textbf{R}}} &
		\multicolumn{1}{c|}{\textit{\textbf{F}}} &
		\multicolumn{1}{c|}{\textit{\textbf{Kappa}}} &
		\multicolumn{1}{c|}{\textit{\textbf{MCC}}} \\ \hline		
		Real + RF      & 0.517                           & 0.909 & 0.531 & 0.063 & \multicolumn{1}{l|}{0.165} \\ \hline
		SMOTE + RF      & 0.721                           & 0.941 & 0.794 & 0.588 & \multicolumn{1}{l|}{0.624} \\ \hline
		CGAN + RF       & \textbf{0.875}       & \textbf{0.942} & \textbf{0.905} & \textbf{0.810} & \multicolumn{1}{l|}{\textbf{0.813}} \\ \hline \hline
		Real + XGBoost & 0.614                           & 0.941 & 0.681 & 0.361 & \multicolumn{1}{l|}{0.448} \\ \hline
		SMOTE + XGBoost & 0.503                           & 0.924 & 0.497 & 0.013 & \multicolumn{1}{l|}{0.075} \\ \hline
		CGAN + XGBoost  & \textbf{0.999} &  \textbf{0.942} & \textbf{0.968} & \textbf{0.937} & \multicolumn{1}{l|}{\textbf{0.939}} \\ \hline \hline
		Real + MLP     & 0.502                           & 0.943 & 0.490 & 0.008 & \multicolumn{1}{l|}{0.062} \\ \hline
		SMOTE + MLP     & 0.506                           & 0.959 & 0.506 & 0.021 & \multicolumn{1}{l|}{0.101} \\ \hline
		CGAN + MLP      & \textbf{0.917}         & \textbf{0.941} & \textbf{0.928} & \textbf{0.857} & \multicolumn{1}{l|}{\textbf{0.857}} \\ \hline	 \hline				
		Real + 1DCNN   & 0.509                           & 0.964 & 0.514 & 0.034 & \multicolumn{1}{l|}{0.128} \\ \hline
		SMOTE + 1DCNN   & 0.505                           & 0.931 & 0.505 & 0.020 & \multicolumn{1}{l|}{0.095} \\ \hline
		CGAN + 1DCNN    & \textbf{0.968} & \textbf{0.941} & \textbf{0.954} & \textbf{0.909} & \multicolumn{1}{l|}{\textbf{0.909}} \\ \hline
	\end{tabular}
\end{table}

Scenario 2 with 865 original samples performed well on RF and XGBoost whereas it fails for neural network models. Irrespective of the imbalance issue and the augmentation methods, RF and XGBoost give satisfactory results. As our focus is on deep learning models, training the model using CGAN merged training data shows a dramatic progress in the metrics. False-positives reduced considerably for MLP and 1DCNN where MLP performs almost similar to RF and XGBoost. The results for scenario 2 experiments are shown in Table \ref{tab:compares2}.

\begin{table}[!hbtp]
	\centering
	\caption{Performance analysis for Scenario 2 Binary Classification}
	\label{tab:compares2}
	\begin{tabular}{|l|l|l|l|l|l|}
		\hline
		\multicolumn{6}{|c|}{\textbf{Scenario 2}}\\ \hline
		\textit{\textbf{Training Models}} &
		\multicolumn{1}{c|}{\textit{\textbf{P}}} &
		\multicolumn{1}{c|}{\textit{\textbf{R}}} &
		\multicolumn{1}{c|}{\textit{\textbf{F}}} &
		\multicolumn{1}{c|}{\textit{\textbf{Kappa}}} &
		\multicolumn{1}{c|}{\textit{\textbf{MCC}}} \\ \hline
		Real + RF       & {0.955}        & {0.918}        & {0.936}        & {0.872}        & 
		{0.873} \\ \hline
		SMOTE + RF      & 0.869          & 0.918          & 0.785          & 0.784          & 0.786          \\ \hline
		CGAN + RF        & \textbf{0.958} & \textbf{0.947} & \textbf{0.952} & \textbf{0.905} & \textbf{0.906} \\ \hline \hline
		Real + XGBoost  & {0.999}        & {0.951}        & {0.974}        & {0.947}         & 
		{0.949} \\ \hline
		SMOTE + XGBoost & 0.723          & 0.995          & 0.806          & 0.614          & 0.664          \\ \hline
		CGAN + XGBoost   & \textbf{0.999} & \textbf{0.870} & \textbf{0.925} & \textbf{0.849} & \textbf{0.859} \\ \hline \hline			
		Real + MLP      & 0.578          & 0.990          & 0.632          & 0.269          & 0.393          \\ \hline		
		SMOTE + MLP     & 0.868          & 0.994          & 0.922          & 0.843          & 0.852          \\ \hline
		CGAN + MLP       & \textbf{0.962} & \textbf{0.988} & \textbf{0.974} & \textbf{0.949} & \textbf{0.949} \\ \hline \hline				
		Real + 1DCNN    & 0.689          & 0.994          & 0.773          & 0.548          & 0.613          \\ \hline
		SMOTE + 1DCNN   & 0.883          & 0.996          & 0.933          & 0.865          & 0.873          \\ \hline
		CGAN + 1DCNN     & \textbf{0.882} & \textbf{0.988} & \textbf{0.928} & \textbf{0.856} & \textbf{0.863} \\ \hline
	\end{tabular}
\end{table}

Extreme imbalance is present in Scenario 3 data with only 20 malicious instances in the dataset. RF and XGBoost give a promising result for such a skewed data. 1DCNN increased its efficacy on boosting the training dataset using CGAN with notable enhancement in precision, F-score and substantial increase in Kappa score and MCC. Table \ref{tab:compares3} provides the performance analysis for scenario 3.

\begin{table}[!htbp]
	\centering
	\caption{Performance analysis for Scenario 3 Binary Classification}
	\label{tab:compares3}
	\begin{tabular}{|l|l|l|l|l|l}
		\hline
		\multicolumn{6}{|c|}{\textbf{Scenario 3}} \\ \hline
		\textit{\textbf{Training Models}} &
		\multicolumn{1}{c|}{\textit{\textbf{P}}} &
		\multicolumn{1}{c|}{\textit{\textbf{R}}} &
		\multicolumn{1}{c|}{\textit{\textbf{F}}} &
		\textit{\textbf{Kappa}} &
		\multicolumn{1}{l|}{\textit{\textbf{MCC}}} \\ \hline
		Real + RF       & {0.749}        & {0.625}        & {0.667}        & {0.333} & \multicolumn{1}{l|}{\textbf{0.333}} \\ \hline
		SMOTE + RF      & 0.536          & 0.625          & 0.555          & 0.011          & \multicolumn{1}{l|}{0.077}          \\ \hline
		GAN + RF        & \textbf{0.833} & \textbf{0.749} & \textbf{0.786} & \textbf{0.571} & \multicolumn{1}{l|}{\textbf{0.577}} \\ \hline \hline					
		Real + XGBoost  & {0.611}        & {0.749}        & {0.654}        & {0.308}        & \multicolumn{1}{l|}{\textbf{0.333}} \\ \hline
		SMOTE + XGBoost & 0.503          & 0.995          & 0.503          & 0.111          & \multicolumn{1}{l|}{0.133}          \\ \hline
		GAN + XGBoost   & \textbf{0.999} & \textbf{0.767} & \textbf{0.833} & \textbf{0.667} & \multicolumn{1}{l|}{\textbf{0.707}} \\ \hline \hline	
		Real + MLP      & 0.500          & 0.984          & 0.495          & 0.004          & \multicolumn{1}{l|}{0.042}          \\ \hline
		SMOTE + MLP     & 0.687          & 0.874          & 0.745          & 0.499          & \multicolumn{1}{l|}{0.530}          \\ \hline
		GAN + MLP       & \textbf{0.749} & \textbf{0.749} & \textbf{0.749} & \textbf{0.399} & \multicolumn{1}{l|}{\textbf{0.452}}          \\ \hline \hline
		Real + 1DCNN    & 0.500          & 0.945          & 0.472          & 0.001          & \multicolumn{1}{l|}{0.022}          \\ \hline
		SMOTE + 1DCNN   & 0.625          & 0.749          & 0.667          & 0.333          & \multicolumn{1}{l|}{0.353}          \\ \hline
		GAN + 1DCNN     & \textbf{0.749} & \textbf{0.875} & \textbf{0.799} & \textbf{0.667} & \multicolumn{1}{l|}{\textbf{0.707}} \\ \hline
	\end{tabular}
\end{table}

The experiments show that the CGAN based training for multi-class classification works as expected for anomaly detection of insider activities. There are a few key findings from these results. The method cannot be tested and proved to be 100\% precise just because of one set of features or one set of data. Since the principal interest of the research was on CGAN based data augmentation for diverse and increased data samples, we chose the simplest feature set for validation. This feature set cannot provide a complete context of the behavior which needs to be enhanced. Area Under the Curve (AUC) score found to be very high for experiments and did not provide any useful analysis; hence not considered in the evaluation. Improved and powerful performance analysis strategies should be used to evaluate and compare learning models. 

The literature shows numerous solution approaches being used in insider threat attack which makes the comparison of methods challenging in terms of data, methods, and the goals. We have tried to incorporate different contexts in which the insider detection has been studied previously and provided compatible comparisons. We selected some results from previous works for a general comparison. The data augmentation strategy used, dataset for evaluation and the learning model are the criteria used for the selection of methods summarized in Table \ref{tab:compare}.

\begin{table}[!htbp]
	\centering
	\caption{Summary of Existing Methods}
	\label{tab:compare}
	\resizebox{\columnwidth}{!}{%
		\begin{tabular}{lllll}
			\hline
			Method & Augmentation & Dataset & Model & No. of Classes \\ \hline
			\cite{ieee_trans_scenario} & Random Over Sampling & CERT 4.2 & \begin{tabular}[c]{@{}l@{}}Deep Autoencoder\\ Random Forest\end{tabular} & Binary \\
			\cite{employee_prof} & \begin{tabular}[c]{@{}l@{}}Create\\ Synthetic Data\end{tabular} & Enron email & Isolation Forest & Binary \\
			\cite{resample_weka} & Spread Subsample & \begin{tabular}[c]{@{}l@{}}Dataset from a \\ market leader\end{tabular} & \begin{tabular}[c]{@{}l@{}}J48 Decision Tree\\ Naive Bayes\\ Random Forest\\ Support Vector Machines\end{tabular} & Binary \\
			\cite{smote_xgb} & \begin{tabular}[c]{@{}l@{}}Data Adjustment \\ with SMOTE\end{tabular} & CERT 6.2 & \begin{tabular}[c]{@{}l@{}}Random Forest\\ Gradient Boosting\\ XGBoost\end{tabular} & Binary \\
			\cite{gan_insider} & VanillaGAN & CERT 4.2 & \begin{tabular}[c]{@{}l@{}}Decision Tree\\ XGBoost\end{tabular} & Binary \\
			Proposed work & Conditional GAN & CERT 4.2 & \begin{tabular}[c]{@{}l@{}}Random Forest\\ XGBoost\\ Neural Network\\ 1DCNN\end{tabular} & Multi-Class \\ \hline
		\end{tabular}%
	}
\end{table}

The scenario-based classification in the work \cite{ieee_trans_scenario} considered the problem as three different binary classification problems. ROS is used for data augmentation under various sampling ratios. The method gave best results for Scenario 2 on RF whereas it could not perform satisfactorily on Scenarios 1 and 2. But our experiments on scenario-based binary classification using CGAN data augmentation gives better results for each scenario. An ensemble strategy combined with data adjusted XGBoost model for insider threat detection proposed in \cite{smote_xgb} gave promising results with an average recall of 98.5\% which is similar to our binary classification results.

Similarly, in the paper \cite{gan_insider}, all scenarios resulted in outstanding performance with almost 100\% for all metrics using GANs. Our experiments using CGAN and scenario-based binary classification gave slightly different results for all scenarios. It can be attributed to many reasons like feature space, GAN architecture and the training. In both the cases, XGBoost performs well as it did in other existing works without any data augmentation\cite{ieee_trans_granular}. 

The method followed in \cite{employee_prof} based on the email communications create a new set of data, Enron\textsuperscript{+}, followed by anomaly detection using Isolation Forest algorithm. Spread subsample, a method that selects the random samples to fit in the memory by balancing the skewed class distributions is mentioned in the paper \cite{resample_weka}. These two works are not comparable in terms of metrics or methods, but provides insight into the possible data re-sampling approaches. In the next section, we summarize the work by discussing the conclusion and scope for future enhancements.

\section{Conclusion}\label{sec_conclusion}
With the aim to improve the detection of malicious insider activities, and to curtail the negative effects of imbalanced class distribution in data, this paper proposed a novel conditional GAN data augmentation followed by anomaly detection for insider activity detection. The proposed method is an end-to-end workflow starting with data pre-processing of the heterogeneous user access log files, feature space creation and feature set generation. The prime focus is on the design of a CGAN for the malicious insider scenarios and the synthetic data generation. Synthetic data quality is analysed using visual methods like kernel density estimation, PCA and t-SNE. CGAN-based data augmentation helps generate more data from original data distribution. Anomaly detection is performed using multi-class classification with four class labels. The performance of proposed approach is evaluated on the benchmark CMU CERT dataset using ensemble and deep learning models. In this work, effectiveness of data augmentation is studied in the context of insider threat analysis. Future work will investigate how GAN performs on different insider threat datasets available and under various feature spaces. Moreover, there are scopes for detailed analysis on the GAN evaluation and performance.


%
%



%

\end{document}